\documentclass[12pt]{iopart}

\usepackage[colorlinks,linkcolor=blue,urlcolor=blue,citecolor=blue,pdfusetitle]{hyperref}
\usepackage[utf8]{inputenc}
\usepackage[english]{babel}
\usepackage[caption = false]{subfig}
\usepackage{graphicx,epstopdf}
\usepackage{blindtext}
\usepackage[table,xcdraw]{xcolor}
\usepackage{lipsum}
\usepackage{amsfonts}
\usepackage{bbm}
\usepackage{amssymb}
\usepackage{color}
\usepackage{latexsym}
\usepackage{times,txfonts}
\usepackage{dsfont}
\usepackage[normalem]{ulem}

\newcommand{\Fcal}{\mathcal{F}}

\newcommand{\1}{\mathbbm{1}}

\newcommand{\ket}[1]{| #1 \rangle}

\newcommand{\eqref}[1]{(\ref{#1})}
\newcommand{\bra}[1]{\langle #1 |}
\newcommand{\ketbra}[2]{| #1 \rangle\langle #2 |}
\newcommand{\abs}[1]{| #1 |}

\newcommand{\SubFig}[2]{\ref{#1}{\color{blue}#2}}

\definecolor{bluePoli}{cmyk}{0.4,0.1,0,0.4}
\definecolor{blueGreen}{RGB}{0, 102, 102}
\definecolor{brickred}{rgb}{0.8, 0.25, 0.33}
\definecolor{darkred}{RGB}{204, 0, 0}
\definecolor{darkgreen}{RGB}{0, 102, 50}
\definecolor{darkblue}{RGB}{0, 76, 153}
\definecolor{mygray}{RGB}{224, 224, 224}

\newcommand{\revisionRefA}[1]{{\color{black}#1}}
\newcommand{\revisionRefB}[1]{{\color{black}#1}}

\newcommand{\UFF}{Instituto de F\'{i}sica, Universidade Federal Fluminense, Av. Gal. Milton Tavares de Souza s/n, Gragoat\'{a}, 24210-346 Niter\'{o}i, Rio de Janeiro, Brazil}
\newcommand{\CSIC}{Instituto de Física Fundamental (IFF), Consejo Superior de Investigaciones Científicas (CSIC), Calle Serrano 113b, 28006 Madrid, Spain}

\usepackage{orcidlink}

\begin{document}
	
	\title[Quantum battery supercharging via counter-diabatic dynamics]{Quantum battery supercharging via counter-diabatic dynamics}

	\author{L. F. C. de Moraes}
	\vspace{-0.2cm}
	\ead{lfcmoraes@id.uff.br}
	\address{\UFF}
	
	\vspace{-0.4cm}
	
	\author{Alan C. Duriez~\orcidlink{0000-0001-5196-0827}}
	\vspace{-0.2cm}
	\ead{alanduriez@id.uff.br}
	\address{\UFF}
	
	\vspace{-0.4cm}
	
	\author{A. Saguia~\orcidlink{0000-0003-0403-4358}}
	\vspace{-0.2cm}
	\address{\UFF}
	
	\vspace{-0.4cm}
	
	\author{Alan C. Santos~\orcidlink{0000-0002-6989-7958}}
	\vspace{-0.2cm}
	\address{*Corresponding author}
	\vspace{-0.2cm}
	\ead{\mailto{ac\_santos@iff.csic.es}}
	\address{\CSIC}
	
	\vspace{-0.4cm}
	
	\author{M. S. Sarandy~\orcidlink{0000-0003-0910-4407}}
	\vspace{-0.2cm}
	\address{\UFF}

	\begin{abstract}
		We introduce a counter-diabatic approach for deriving Hamiltonians modeling superchargable quantum batteries (QBs). A necessary requirement for the supercharging process is the existence of multipartite interactions among the cells of the battery. Remarkably, this condition may be insufficient no matter the number of multipartite terms in the Hamiltonian. We analytically illustrate this kind of insufficiency through a model of QB based on the adiabatic version for the Grover search problem. On the other hand, we provide QB supercharging with just a mild number of global connections in the system. To this aim, we consider a spin-$1/2$ chain with $n$ sites in the presence of Ising multipartite interactions. We then show that, by considering the validity of the adiabatic approximation and by adding $n$ terms of $(n-1)$-site interactions, we can achieve a Hamiltonian exhibiting maximum QB power, with respect to a normalized evolution time, growing quadratically with $n$. Therefore, supercharging can be achieved by $O(n)$ terms of multipartite connections. The time constraint required by the adiabatic approximation can be surpassed by considering a counter-diabatic expansion in terms of the gauge potential for the original Hamiltonian, with a limited $O(n)$ many-body interaction terms assured via a Floquet approach for the counter-diabatic implementation.    
	\end{abstract}
	
	\maketitle
	\section{Introduction}
	\label{intro}
	
	Quantum batteries (QBs) are devices able to store and distribute energy in a quantum network~\cite{Alicki:13,PRL2013Huber,Campaioli:Book,PRL_Andolina}. They are fundamental ingredients for quantum processes for which energy transfer is triggered by specific events, being stored otherwise. Examples are provided by quantum thermal machines~\cite{kieu:04,Bhattacharjee:21}, where energy should be provided to and extracted from a quantum system in parts of a cycle. In order to be applied in a real scenario, QBs are expected to exhibit several features. Firstly, it is desired that QBs behave as {\it reliable} devices, keeping its charge under parameter fluctuations and environment effects (see, e.g., Refs.~\cite{Campaioli:23,PRB2019Batteries,Carrega:20,Mojaveri:23,Arjmandi:23,Lu:24,Ahmadi:24}). Secondly, QBs must be {\it stable} as a function of time, which means that the dynamics asymptotically drive the system to a charged or discharged state. The stability problem can be solved by an adiabatic quantum evolution~\cite{Santos:19-a,Santos:20PRE,Moraes:21,Abel:24}. A third desired property, which radically distinguishes QBs from classical devices, is the supercharging behavior~\cite{Alicki:13}. Indeed, it has been shown that, due to quantum correlations~\cite{gyhm_beneficial_2024}, QBs may attain a fast charging regime, with maximum power growing quadratically with respect to the size $n$ of the quantum system~\cite{Celeri:17PRL,Rossini:20-PRL,Gyhm:22}, while classical (non-interacting) batteries exhibit power behaving linearly with $n$. In this supercharging scenario, the quantum advantage, as measured by the ratio between the maximum quantum and classical power, shows then an extensive scaling with respect to the size $n$ of the system.
	
	The supercharging property requires, as a necessary but not sufficient condition, that multipartite (global) interactions are present in the quantum system~\cite{Gyhm:22}. Remarkably, what provides sufficiency for supercharging is still under debate. Here, we introduce a counter-diabatic approach for deriving Hamiltonians modeling superchargable QBs. We start by analytically illustrating the insufficiency of global couplings through a model of QB based on the adiabatic version for the Grover search problem~\cite{Roland:02}. Our aim with this example is to corroborate the idea that the underlying mechanism behind supercharging is a nontrivial problem. On the other hand, we provide an adiabatic scheme for supercharging that may involve just a mild number of multipartite connections in the system. In this direction, we consider a spin-$1/2$ chain 
	with $n$ sites in the presence of Ising multipartite interactions\revisionRefB{, which constitutes a model of spin QB~\cite{Le:18,Barra_2022,Grazi:24,Salvia:23,Santos:21,Dou:22}.} 
	In this model, the time-dependent Hamitonian $H(\lambda(s))$ depends on a smooth bounded function $\lambda(s)$, which is defined in terms of a normalized parameter $s=t/T$ ($0 \le s \le 1$), with $t$ denoting the instantaneous time and $\tau$ the total time of evolution. We then show that, by considering the validity of the adiabatic approximation and by adding $n$ terms of 
	$(n-1)$-site interactions, we can achieve a Hamiltonian exhibiting maximum QB power, with respect to the normalized time $s$, 
	growing quadratically with $n$. 
	Therefore, supercharging can be achieved by $O(n)$ terms of multipartite connections. 
	Naturally, as we increase $n$, the power with respect to the instantaneous time $t$ will be affected by the adiabatic time constraint, potentially reducing the power performance. However, the detrimental effects of the total adiabatic time, which increases with the size of the system, can be surpassed by considering a counter-diabatic driving for the original Hamiltonian~\cite{Demirplak:03,Demirplak:05,Berry:09}. As usual, the exact counter-diabatic dynamics remove the adiabaticity constraints over the evolution time at the expense of adding non-local terms in the total Hamiltonian~\cite{Torrontegui:13}. We solve this problem by considering a counter-diabatic series for the gauge potential~\cite{KOLODRUBETZ20171}, which approximates the adiabatic dynamics at arbitrary precision. The expansion can be implemented through the original $O(n)$ global terms in the Hamiltonian within a Floquet approach~\cite{Claeys:19}. Therefore, as we will show, we can keep supercharging in the Floquet Ising QB with just a mild set of $O(n)$ multipartite interactions, which may be fully implemented with current technology.

	\section{Grover Battery}
	\label{sec:sec1}
	
	In this section, we introduce a charging process of a QB based on the adiabatic realization of the Grover algorithm~\cite{Roland:02}, where the desired ("marked") element is an excited state of the battery. For the sake of comparison, we first introduce the driving Hamiltonian for a classical (parallel) charging of single-qubit cells of a QB as 
	\begin{equation}
		H_{\parallel}(t) = \Omega_0\sum_{k=1}^{n}\left[f(t) \sigma^{(k)}_x + g(t)\sigma^{(k)}_z\right] = \sum_{k=1}^{n}H_{\parallel}^{k}(t), \label{eq:LZ}
	\end{equation} 
	where $\Omega_0$ sets the Hamiltonian energy scale, $n$ is the number of cells of the battery, and $\sigma_x$ and $\sigma_z$ are the Pauli operators defined in the computational basis $\{\ket{n}\}$, with $n \in \{0,1\}$, such that $\sigma_z \ket{n} = (-1)^{n+1} \ket{n}$, $\sigma_x \ket{0} = \ket{1}$, and $\sigma_x \ket{1} = \ket{0}$. 
	Throughout the manuscript, we adopt $\hbar = 1$. As it can be seen, the above Hamiltonian drives a system of $n$ quantum cells (e.g., spin$-1/2$ particles) in a parallel (non-interacting) charging, where the $k$-th cell is independently driven by the local Hamiltonian $H_{\parallel}^{k}(t)$. 
	
	The system ergotropy~\cite{Allahverdyan:04}, {\it i.e.}, the extractable energy by unitary operation, is defined here with respect to the reference Hamiltonian $H_{\mathrm{ref}} = \sum_{k} \Omega_0 \sigma^{(k)}_z$. Therefore, by choosing the time-dependent functions $g(t)$ and $f(t)$ such that $g(0) = f(\tau) = 1$ and $g(\tau) = f(0) = 0$, for a sufficiently large time $\tau$, the $k$-th quantum cell will be driven from the ground state of its Hamiltonian $ \Omega_0 \sigma^{(k)}_z$ to the ground state of the Hamiltonian $ \Omega_0\sigma^{(k)}_x$, which is a charged state with respect to the reference Hamiltonian $H_{\mathrm{ref}}$. Globally, according to the adiabatic theorem, the system driven by the Hamiltonian $H_{\parallel}(t)$ will evolve from the initial state $\ket{E_{\mathrm{empty}}}=\ket{0}^{\otimes n}$ (passive  state of $H_{\mathrm{ref}}$), to the half-charged state $\ket{E_{\mathrm{hc}}} = \ket{+}^{\otimes n}$, with $\ket{+} = (\ket{0} +\ket{1})/\sqrt{2}$. \revisionRefB{Notice that the structure of the Hamiltonian in Eq.~\eqref{eq:LZ} and the boundary conditions of the driving functions $f(t)$ and $g(t)$ do not allow the fully-charged state $\ket{E_{\mathrm{full}}}=\ket{1}^{\otimes n}$ to be reached at the end of the protocol. Therefore, for any choice of $f(t)$ and $g(t)$ satisfying the aforementioned boundary conditions, the half-charged state $\ket{E_{\mathrm{hc}}}$ is indeed the maximally charged state we can achieve by driving the battery through our Hamiltonian}.
	Since we are transitioning throughout pure states, the ergotropy available for extraction is then provided by the  difference $\Delta E = E_{\mathrm{hc}}-E_{\mathrm{empty}} = n  \Omega_0$ between the energy $E_{\mathrm{hc}} = \langle E_{\mathrm{hc}} |H_{\mathrm{ref}} | E_{\mathrm{hc}} \rangle = 0$ of the half-charged state and the energy $E_{\mathrm{empty}}=\langle E_{\mathrm{empty}} |H_{\mathrm{ref}} | E_{\mathrm{empty}} \rangle = - n \Omega_0$ of the empty state. This scenario constitutes the classical charging scheme to be used as a benchmark for the study of the collective (quantum) charging protocols.
	
	Let us now introduce the Grover battery, which is a model for QB with global interactions based on the adiabatic realization of the Grover algorithm~\cite{Roland:02}. By considering a Hilbert space of dimension $N=2^n$, we define the driving Hamiltonian by 
	\begin{equation}
		H(s) = f(s)(\1-\ketbra{E_{\mathrm{hc}}}{E_{\mathrm{hc}}})+g(s)(\1-\ketbra{E_{\mathrm{empty}}}{E_{\mathrm{empty}}}) ,
		\label{eq:grover_hamiltonian}
	\end{equation}
	where we have defined the normalized time $s=t/\tau$. 
	By preparing the system in the empty energy state $\ket{E_{\mathrm{empty}}}$ and by letting the system evolve under adiabatic dynamics from $t=0$ until $t=\tau$, this Hamiltonian allows us to achieve the same final state as the parallel Hamiltonian $H_{\parallel}(t)$, but now through  collective (global) interactions. Notice that the final (target) state can be expressed as
	\begin{equation}
		\ket{E_{\mathrm{hc}}} = \ket{+}^{\otimes n} = \frac{1}{\sqrt{N}}\sum_{i=0}^{N-1}\ket{i} , \label{eq:initial_state}
	\end{equation}
	which is a uniform superposition of computational basis states\revisionRefA{, as denoted by the set $\{ \ket{i} \mid i \in \{0, \ldots, N-1\} \}$}. Therefore, the system evolves from the discharged state $\ket{E_{\mathrm{empty}}}$ to the half-charged (target) state $\ket{E_{\mathrm{hc}}}$ of the battery\revisionRefB{, which is the maximally charged state attainable in our protocol, as aforementioned}.
	
	\revisionRefA{By a quick inspection of the Hamiltonian in Eq.\eqref{eq:grover_hamiltonian}, we can see that every state orthogonal to both $\ket{\psi_0} \equiv \ket{0}$ and $\ket{\psi_1} \equiv \frac{1}{\sqrt{N-1}}\sum_{i=1}^{N-1}\ket{i}$ is a trivial eigenvector of $H(s)$, with unity eigenvalue. This degenerate subspace is isolated from the dynamics by choosing an initial state in the subspace spanned by $\{\ket{\psi_0},\ket{\psi_1}\}$, which is the case for $\ket{E_{\mathrm{empty}}} = \ket{\psi_0}$. By considering this property, we can describe the dynamics by only taking two independent complex amplitudes, with the state vector reading}
	\begin{equation}
		\ket{\psi(s)} = a(s)\left[ \ket{0}+k(s)\sum_{i=1}^{N-1}\ket{i}\right], \label{eq:state-vector}
	\end{equation}
	\revisionRefA{where $a(s)$ is the amplitude of the state $\ket{\psi_0}$ and $k(s)$ is the ratio between the amplitudes of $\ket{\psi_1}$ and $\ket{\psi_0}$.} \revisionRefA{The} normalization of the state vector provides 
	\begin{equation}
		\abs{a(s)}^2 = \frac{1}{1+(N-1)\abs{k(s)}^2} ,\label{eq:norm-condition}
	\end{equation}
	with $k(s)$ determined by the quantum dynamics. \revisionRefB{The role played by $k(s)$ in Eq.~\eqref{eq:state-vector} is to control the weights of $\ket{\psi_1}$ and $\ket{\psi_0}$ in the superposition state $\ket{\psi(s)}$, driving the system towards the target state $\ket{E_{\mathrm{hc}}}$ at the end of the evolution. Notice that, since $a(s)$ can be fixed by the normalization condition, the general solution for the Grover dynamics is then provided by $k(s)$. In the case of an adiabatic evolution, $k(s)$ can be \textit{unequivocally} obtained by imposing $\ket{\psi(s)}$ in Eq.~\eqref{eq:state-vector} as the instantaneous ground state of $H(s)$~\cite{Roland:02,Coulamy:17}.}
	The performance of the QB is governed by its  instantaneous power, which can be defined as 
	\begin{equation}
		P(s)=\frac{1}{\tau}\frac{d}{ds}\bra{\psi(s)}H_{\mathrm{ref}}\ket{\psi(s)}. \label{eq:power_def}
	\end{equation} 
	By using that $H_{\mathrm{ref}}$ is diagonal in the computational basis and that $\Tr[{H_{\mathrm{ref}}}]=0$, we can obtain the instantaneous power in terms of $k(s)$ as given by 
	\begin{equation}
		P(s) = \frac{\Delta E}{\tau}\frac{N}{\left(1+(N-1)\abs{k(s)}^2\right)^2}\frac{d}{ds}\abs{k(s)}^2, \label{eq:instantaneous-power}
	\end{equation}
	where we have explicitly inserted the normalization condition provided by Eq.~\eqref{eq:norm-condition}. From Eq.~\eqref{eq:instantaneous-power}, we can analyze the charging properties of the Grover Hamiltonian. 
	
	%
	We will consider the adiabatic evolution for the QB driven by the Hamiltonian $H(s)$ in Eq.~\eqref{eq:grover_hamiltonian}. This is due to the fact that, as previously mentioned, adiabaticity allows for a stable charging protocol~\cite{Santos:19-a,Santos:20PRE,Moraes:21}. In that case, we can analytically determine the amplitude $k(s)$ for the state vector in Eq.~\eqref{eq:state-vector}. Starting the evolution in the ground state of $H(0)$, the system will continuously evolve to the instantaneous ground state of $H(s)$, with $k(s) = k_{\mathrm{ad}}(s)$ given by
	\begin{equation}
		k_{\mathrm{ad}}(s) = 1-\frac{1-\sqrt{1-4 f(s)g(s)\overline{N}}}{2\overline{N}f(s)},
	\end{equation}
	where $\overline{N}\equiv1-1/N$. \revisionRefA{The above expression for $k_{ad}(s)$ can be obtained by directly imposing that $\ket{\psi(s)}$ in Eq.~\eqref{eq:state-vector} is the adiabatic solution of the dynamics driven by $H(s)$, so that $\ket{\psi(s)}$ satisfies the eigenvalue equation 
		$H(s)\ket{\psi(s)}=E_{\mathrm{gs}}(s)\ket{\psi(s)}$,
		with $E_{\mathrm{gs}}(s)=\left[1-\sqrt{1-4 f(s)g(s)\overline{N}}\right]/2$ denoting the instantaneous ground state energy [as the system is assumed to be initially prepared in the ground state of $H(s=0)$].} 
	
	We consider two different interpolation protocols in Eq.~\eqref{eq:grover_hamiltonian}, the linear interpolation, defined as
	\begin{equation}
		f_\mathrm{lin}(s) = s , ~~ g_\mathrm{lin}(s) = 1-s,
	\end{equation}
	and the optimal interpolation~\cite{Roland:02}, as given by quantum adiabatic brachistochrone~\cite{Rezakhani:09,Santos:21}
	\begin{equation}
		f_\mathrm{Brach}(s) = \frac{\sqrt{N-1}-\tan\biggl[(1-2s)\arctan\Big(\sqrt{N-1}\Big)\biggr]}{2\sqrt{N-1}} ,
	\end{equation}
	\begin{figure}[t]\centering
		\includegraphics[width=\linewidth]{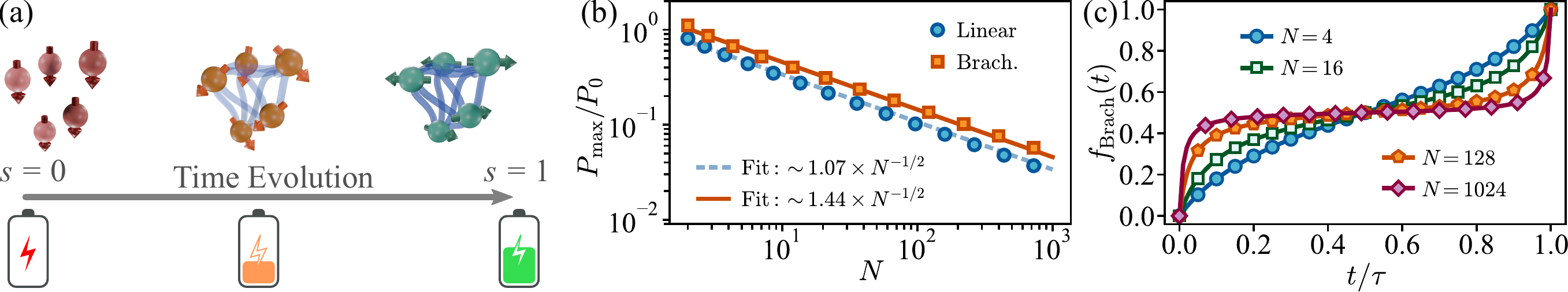}
		\caption{(a) Schematic description of the charging protocol, with the QB initially at the empty charge state. Throughout the quantum evolution, ergotropy is injected in the QB until its final half-charged state. (b) Scaling of the maximum power for both linear and optimal interpolations. 
			We have added the function $N^{-1/2}$ as a dashed line for comparison. We have also  normalized $P_{\mathrm{max}}$ by defining $P_{0} = \Omega_{0}\Delta E$. (c) Behavior of the brachistochrone as function of the time for different QB sizes.
		}\label{fig:power_scaling_ad}
	\end{figure}
	and $g_\mathrm{Brach}(s) = 1-f_\mathrm{Brach}(s)$. Notice that the charging process through the Grover Hamiltonian in  Eq.~\eqref{eq:grover_hamiltonian} requires the emergence of many-body interacting terms, including $n$-body contributions (as shown in Fig.~\SubFig{fig:power_scaling_ad}{a}).
	
	Similarly to the adiabatic realization of the Grover algorithm, the total charging time scales as $\tau_\mathrm{lin} \rightarrow O(N)$ for the linear interpolation  and $\tau_\mathrm{Brach} \rightarrow O(\sqrt{N})$ for the optimal interpolation~\cite{Roland:02}. By taking into account each interpolation schedule and the expression for $k_{\mathrm{ad}}(s)$, we can obtain the instantaneous power from Eq.~\eqref{eq:instantaneous-power},  
	where we consider the total evolution time $\tau$ as $\tau_\mathrm{lin} = N/\Omega_{0} $ and $\tau_\mathrm{Brach} =  \sqrt{N}/\Omega_{0}$.
	Then, we can investigate the scaling of the charging power as a function of $N$. \revisionRefB{As a metric of performance for the QB, we will take the behavior of the maximum \textit{instantaneous} power throughout the evolution, which is defined by $P_{\mathrm{max}} = \max_s{P(s)}$. This adoption is mainly motivated by the metric introduced in  Ref.~\cite{Gyhm:22} to prove the necessity (but no sufficiency) of global interactions to achieve the supercharging behavior.}
	In Fig.~\SubFig{fig:power_scaling_ad}{b}, we show the scaling of the maximum power as a function of $N$ for the two interpolations considered. 
	Notice that, for both interpolation protocols, $P_{\mathrm{max}}$ decreases with the number of qubits approximately as $N^{-1/2}=2^{\,(-n/2)}$, failing to achieve the classical power associated with independent batteries in parallel, which grows linearly with $n$. Remarkably, the scaling of the brachistochrone achieves a slightly enhancement with respect to the linear implementation. This result has been previously observed, for the case of  single-cells superconducting QBs~\cite{Hu:22a}. Our model suggests the generalization of such a result to many-cells QBs. Here, this advantage comes mainly as a response of the change of the shape of the brachistochrone curve as $N$ increases, as shown in Fig.~\SubFig{fig:power_scaling_ad}{c}. In fact, notice that the asymptotic flat shape arises due to the brachistochrone curve sensitivity to changes in the minimum energy gap of the Grover Hamiltonian, around $s = 0.5$, for different values $N$.
	
	We could accelerate the adiabatic algorithm by adding counter-diabatic fields and interactions to the Hamiltonian~\cite{Demirplak:03,Demirplak:05,Berry:09}. However, this will, at most, remove the adiabatic time constraint, multiplying $P_{\mathrm{max}}$ by the total time $\tau$ of evolution, which behaves as $O(N)$ for the linear interpolation and $O(\sqrt{N})$ for the optimal case. Therefore, even by adopting a counter-diabatic dynamics, the power $P_{\mathrm{max}}$ still fails to achieve a classical power behavior. This illustrates the insufficiency of many-body interactions to achieve QB supercharging.

	\section{Many-body Ising battery in a transverse magnetic field}
	
\begin{figure}[t]\centering
		\includegraphics[width=0.75\linewidth]{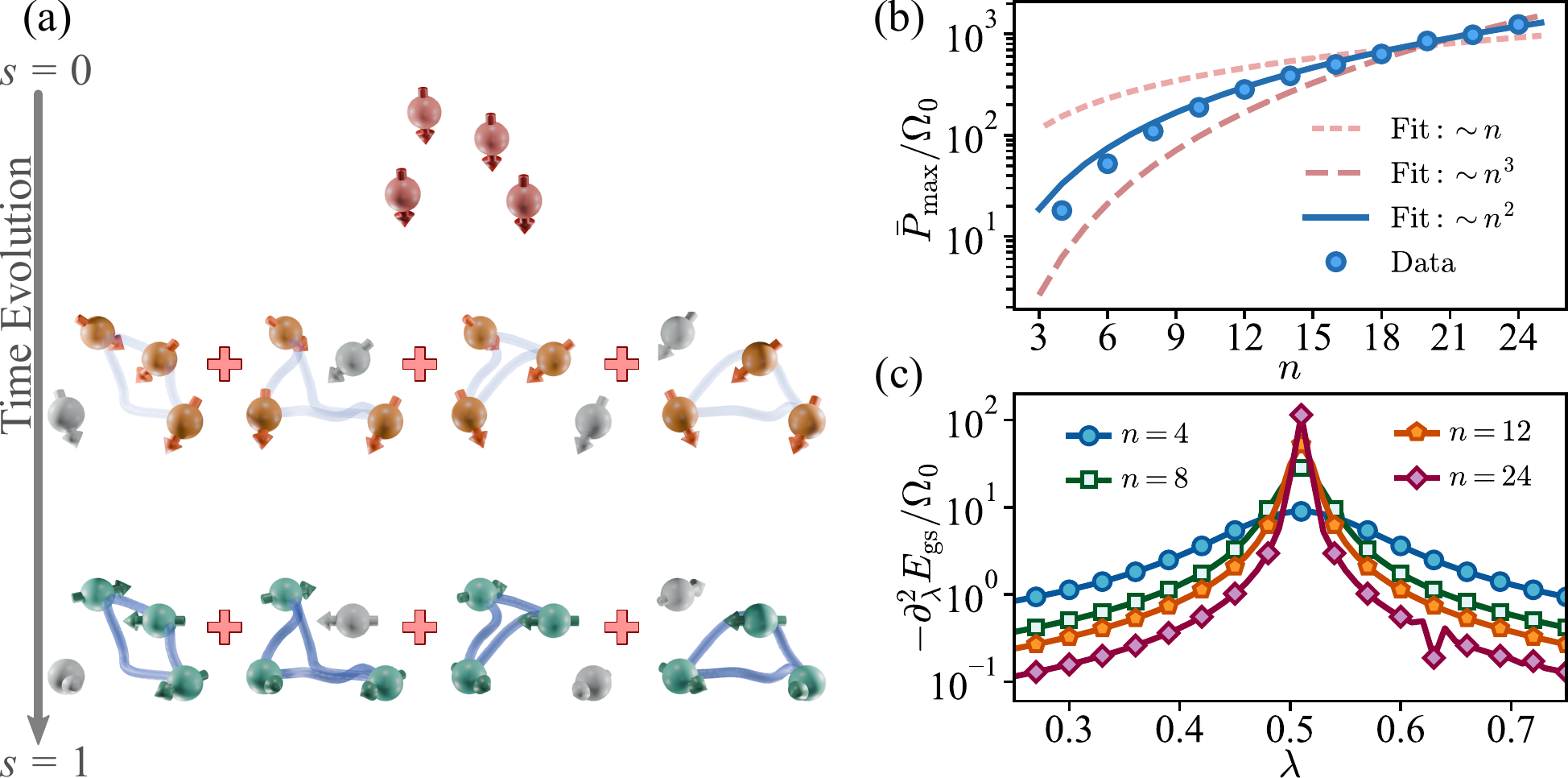}
		\caption{(a) Schematic description of the $(n-1)$-spin interaction terms of the Ising Hamiltonian in Eq.~(\ref{Isingmb}) (for the case $n=4$). Differently from the Grover QB, $n$-spin interaction terms are not present in the system. (b) Scaling of the maximum power for the adiabatic dynamics of the multi-qubit Ising spin-$1/2$ interaction in a magnetic field. (c) The second derivative of the ground state energy with respect to $\lambda$, which shows a second-order (continuous) quantum phase transition in the model.}\label{fig:quad_power}
	\end{figure}

	Let us now discuss a QB model exhibiting optimal supercharging behavior with a just mild set of many-body interactions. For the sake of comparison with the Grover charging, here we will consider the same reference Hamiltonian given by $H_{\mathrm{ref}}$ as before. As depicted in Fig.~\SubFig{fig:quad_power}{a}, differently from the Grover battery, the time-dependent Hamiltonian is provided by a spin-$1/2$ chain with $n$ sites under a $z$-basis local magnetic field supplied by $n$ terms of $(n-1)$-body interactions, reading
	\begin{equation}
		H(\lambda(s)) = (1-\lambda(s)) H_{\mathrm{ref}} + \lambda(s)  \Omega_0 \sum_{i=1}^{n} \prod_{j\ne i}^{n} \sigma^x_j ,
		\label{Isingmb}
	\end{equation}
	where $\lambda(s)$ is an oscillating interpolation function given by 
	\begin{equation}
		\lambda(s) = \sin^2\left[\left(\frac{\pi}{2}\right) \sin^2\left(\frac{\pi\, s}{2}\right)\right] .
	\end{equation}
	Notice that, since the dimensionless time $s$ varies from $0$ to $1$ throughout the evolution, the oscillating parameter $\lambda(s)$ behaves as $0\le \lambda(s) \le 1$. We will then adiabatically evolve the system from the ground state of a pure magnetic field in the $z$ direction (the discharged state) until the ground state of a pure multi-qubit Ising interaction in the $x$ direction (the half-charged state). Indeed, as shown in Ref.~\cite{Gyhm:22}, global interactions are necessary to achieve supercharging behavior, even though they are far from sufficient for this task, as we have  explicitly shown for the Grover battery. Concerning the specific oscillatory form of the driving protocol $\lambda(s)$, it is motivated by the adiabatic optimization of boundary conditions, so that the first and the second derivatives of $\lambda(s)$ with respect to $s$ vanish at the beginning and end of the protocol~\cite{Rezakhani:10,Hu-Biao:16}.
	
	In order to investigate the maximum power for the adiabatic implementation of the QB, we will start by redefining the power to a normalized version with respect to the dimensionless time $s$, which is given by
	\begin{equation}
		\overline{P}(s)= \frac{d}{ds}\bra{\psi(s)}H_{\mathrm{ref}}\ket{\psi(s)} . \label{eq:power_def-Ising}
	\end{equation}
	
	Since we are interested in the scaling performance of the battery with respect to the system size $n$, we now consider the maximum power $\overline{P}_{\mathrm{max}}(n) = \max_s{\overline{P}(s)}$ with respect to $s$. In Fig.~\SubFig{fig:quad_power}{b}, we show the scaling of $\overline{P}_{\mathrm{max}}(n)$ as we increase the number of sites $n$. Notice that $\overline{P}_{\mathrm{max}}(n)$ achieves the quadratic optimal behavior expected for QB, obtaining therefore the quantum extensive advantage with respect to the classical linear behavior.  
	
	In order to achieve supercharging, we must now show that the maximum power with respect to the instantaneous time $t$ (instead of $s$) keeps the quadratic behavior as we increase $n$. By defining the maximum instantaneous power $P_{\mathrm{max}}$ as $P_{\mathrm{max}} = (1/\tau) \overline{P}_{\mathrm{max}}$, we must take into account the scaling of $\tau$ as a function of $n$. Notice that the adiabatic dynamics requires a slow evolution, with a total time $\tau$  constrained to be much greater than the inverse of the minimal energy gap squared~\cite{Messiah:Book}. In order to understand the scaling of the gap, we can look at the second-order derivative of the ground state energy with respect to $\lambda(s)$. This is shown in Fig.~\SubFig{fig:quad_power}{c}. This plot indicates the existence of a second-order (continuous) quantum phase transition for the model, which implies a polynomial shrinking of the gap as we increase the size $n$ of the system. Therefore, the maximum instantaneous power $P(n)$ will be polynomially reduced due to the adiabatic constraint over $\tau$. Naturally, this is detrimental for the supercharging behavior, since the quadratic scaling will be lost in view of the polynomial scaling of $\tau$ with respect to $n$. 
	
	We can restore the supercharging scaling by applying a  counter-diabatic correction to the Hamiltonian. To this end, we will consider the time-dependent Hamiltonian as  $H(\lambda(t))$, with $t=s \tau$, so that we aim at removing the constraint over the total time $\tau$. The auxiliary counter-diabatic terms allow for a fast charging protocol by compensating the diabatic excitations that naturally occur in fast evolutions. For a general Hamiltonian, the counter-diabatic correction can be added so that we obtain a counter-diabatic Hamiltonian $H_{CD}(t)$ in the form
	\begin{equation}
		H_{CD}(t) = H(\lambda(t)) + \dot{\lambda} {\cal{A}}_{\lambda}(t),
	\end{equation}
	where $\dot{\lambda} \equiv d\lambda/dt$ and 
	${\cal{A}}_{\lambda}(t)$ is the adiabatic gauge potential~\cite{KOLODRUBETZ20171}, whose matrix elements read
	\begin{equation}
		\langle m | {\cal{A}}_{\lambda} | n \rangle = 
		i \langle m  | \partial_\lambda n \rangle, 
	\end{equation}
	with $\{|n\rangle\}$ denoting the eigenbasis of the original Hamiltonian $H(\lambda)$. The adiabatic potential can be approximated by an expansion given by~\cite{Claeys:19} 
	\begin{equation}
		{\cal{A}}_{\lambda}^{(\ell)} = i 
		\sum_{k=1}^{\ell} \alpha_{k} \underbrace{\left[H(\lambda),\left[H(\lambda),\cdots,\left[H(\lambda)\right.\right.\right.}_{2k-1},
		\left.\left.\left.\hspace{-0.1cm}\partial_\lambda H(\lambda)\right]\right]\right],
		\label{app-gp}
	\end{equation}
	where $\ell$ determines the order of the expansion and the set $\{\alpha_k\}$ denotes variational parameters to be fixed by minimizing the action $S_\ell$:
	\begin{equation}
		S_\ell = \Tr{G_\ell^2}, \hspace{0.25cm} 
		G_\ell^2 = \partial_\lambda H(\lambda) - 
		i \left[H(\lambda),{\cal{A}}_{\lambda}^{(\ell)}\right]. 
		\hspace{0.5cm}
	\end{equation}
	We observe that the exact gauge potential can be 
	recovered in the limit $\ell \rightarrow \infty$~\cite{Claeys:19}. For a finite $\ell$, the approximate gauge potential can be directly computed from Eq.~(\ref{app-gp}). This approach turns out to entail the inclusion of further many-body interaction terms, which would invalidate the picture of supercharging behavior with just $O(n)$ many-body interaction terms. 
	
	In order to avoid the inclusion of extra global interactions beyond $O(n)$ contributions of $(n-1)$-body terms, we will engineer the gauge potential with a Floquet protocol. The Floquet Hamiltonian can be seen as an oscillating approximation for the counter-diabatic Hamiltonian, reading
	\begin{equation}
		H_\Fcal(t) = \left[1+\frac{\omega}{\omega_0}\cos(\omega t)\right]H(\lambda(t)) + \left[\sum_{k=1}^{\infty} \beta_k \sin[(2k-1)\omega t]\right]\frac{d H}{d \lambda} \frac{d\lambda}{dt}, \hspace{0.7cm} \label{H_shortcut}
	\end{equation}
	where $\omega_0$ is a reference frequency, $\omega$ is the Floquet frequency, and $\beta_k$ are the Fourier coeeficients which, up to second order, are given by~\cite{Claeys:19}
	\begin{equation}
		\beta_1 = 2\omega_0 \alpha_1 , ~~
		\beta_2 = 2\omega_0 \left[24\omega_0^2\alpha_2 + 3 \alpha_1\right] .
	\end{equation}
	The reference frequency $\omega_0$ is 
	typically set by the excitation energy of the system, while the Floquet frequency is taken as much greater than $\omega_0$. Notice also that the Fourier coefficients $\beta_k$ depend on the variational parameters $\alpha_k$ introduced  in Eq.~(\ref{app-gp}) for the approximation of the gauge potential.
	
	\begin{figure}[t]\centering
		\includegraphics[scale=0.25]{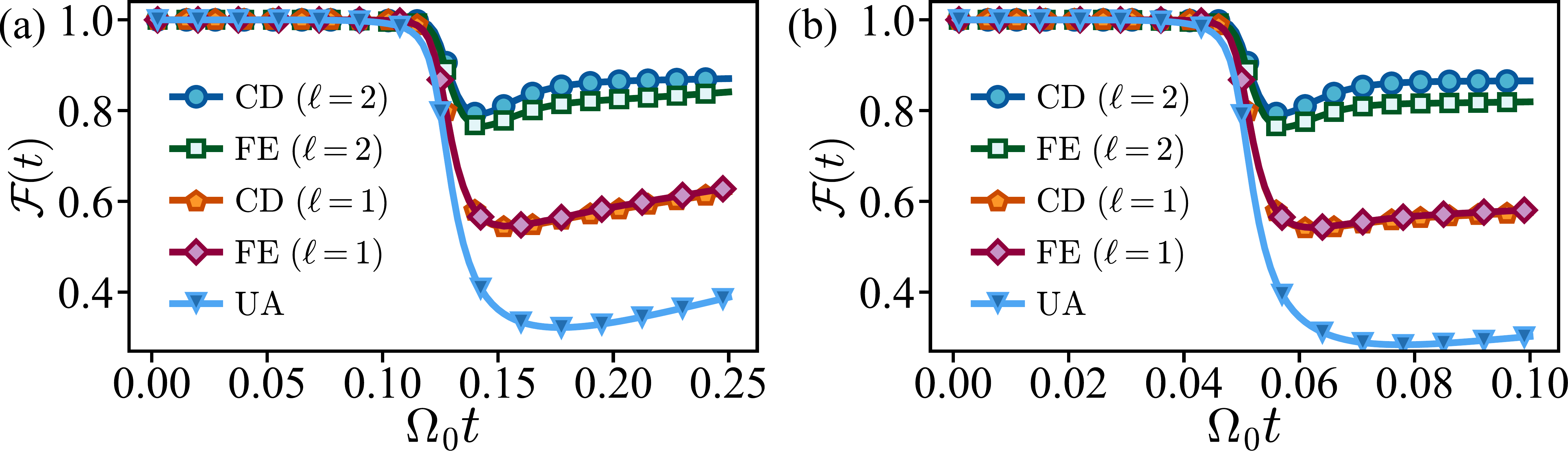}
		\caption{Fidelity $\Fcal(t)$ with respect to the adiabatically evolved state for the unassisted dynamics under $H(\lambda(t))$, the counter-diabatic (CD) expansions for $\ell=1,2$ and the corresponding Floquet evolutions (FE) for $\ell=1,2$. The length of the chain is taken as $n=4$ and the total time of evolution is set as (a) $\tau\Omega_{0}=0.25$ and (b) $\tau\Omega_{0}=0.1$. The frequencies are $\omega_0 = 2 \pi \Omega_{0}$ and $\omega = 10^3 \omega_0$ for $\ell=1$ and $\omega = 10^4\omega_0$ for $\ell=2$.}\label{fig:fid1}
	\end{figure}
	
	As a figure of merit for the battery charging process, we look at the fidelity $\Fcal(t)$ between the dynamically evolved state $|\psi(t)\rangle$ and the adiabatically evolved state $|\psi_{ad}(t)\rangle$, which would drive the system to a charged state. Mathematically, we can write $\Fcal(t) = \left| \langle \psi_{ad}(t) | \psi(t) \rangle\right|$. 
	Now, as an illustration, we set $n=4$ and consider two choices for the total time of evolution: $\Omega_{0}\tau=0.25$ (as in Fig.~\SubFig{fig:fid1}{a}) and $\Omega_{0}\tau=0.1$ (as in Fig.~\SubFig{fig:fid1}{b}), both of which are much less than the adiabatic time scale. Then, in Fig.~\ref{fig:fid1}, we plot the fidelity $\Fcal(t)$ for a number of distinct evolutions, namely, the unassisted evolution (UA) under $H(\lambda(t))$, the counter-diabatic (CD) expansions for $\ell=1,2$ and the corresponding Floquet evolutions (FE) for $\ell=1,2$. Notice that the UA evolution is unable to drive the system to the target state, since we are in a fast charging regime ($\Omega_{0}\tau$ is too small and, therefore, the dynamics is far from the adiabatic approximation) . On the other hand, the CD expansion up to order $\ell=2$ is able to approximate the dynamics to the adiabatic evolution. 
	Remarkably, this approximation to the target state can be also well attained by the second-order Floquet Hamiltonian in Eq.~\eqref{H_shortcut}. \revisionRefB{Concerning experimental feasibility, notice that $H_\Fcal(t)$ is preferable in comparison with the expansion of the gauge potential ${\cal{A}}_{\lambda}(t)$ in Eq.~\eqref{app-gp}. Indeed, a direct  implementation of ${\cal{A}}_{\lambda}(t)$ would require the engineering of global interactions beyond those initially present in the original Hamiltonian, due to the series of commutators  in Eq.~\eqref{app-gp}. 
		By considering the simulation of global interactions through two-qubit gates with tunable couplings, potential platforms to implement the Ising QB include trapped ions~\cite{Leibfried:03} and superconducting integrated circuits~\cite{wendin2017}, as they allow for efficient engineering of time-dependent interactions~\cite{Rasmussen:21,Hu:23,Katz:23}. Moreover, the synthesis of many-body interactions in superconducting quantum circuits has also been put forward~\cite{Zhang_SQC:22}, which can be a tool for achieving QB supercharging.} 
	
	It is worth mentioning that the quantum charging considered in the many-body Ising QB does not violate the requirements of genuine quantum advantage~\cite{Celeri:17PRL,Gyhm:22}. In fact, to identify whether a QB is taking advantage of artificial coupling resources during its charging process, we can compute the ratio between the \textit{driving potential} parameter $v^{\mathrm{dv}}$ for each evolution. Such a parameter is defined for the parallel $v_{\parallel}^{\mathrm{dv}}$ and collective $v_{\#}^{\mathrm{dv}}$ charging as
	\begin{equation}
		v_{\parallel}^{\mathrm{dv}} = ||\hat{V}_{\parallel} - v^{\mathrm{min}}_{\parallel}|| , \quad v_{\#}^{\mathrm{dv}} = ||\hat{V}_{\#} - v^{\mathrm{min}}_{\#}|| .
	\end{equation}
	where $||\cdot||$ denotes operator norm and $\hat{V}_{\parallel}$ and $\hat{V}_{\#}$ are the charging potential for the parallel and collective charging Hamiltonians, respectively, with $v^{\mathrm{min}}_{\parallel}$ and $v^{\mathrm{min}}_{\#}$ their minimum eigenvalue. As observed in Refs.~\cite{Celeri:17PRL,Gyhm:22}, whenever the ratio $r = v_{\#}^{\mathrm{dv}}/v_{\parallel}^{\mathrm{dv}} \le 1$, the QB is not taking advantage of artificial energy resources during its charging performance. Therefore, by taking the Hamiltonian in Eq.~\eqref{eq:LZ} as the parallel charging and the Hamiltonian in Eq.~\eqref{Isingmb} as the collective charging, we identify that $\hat{V}_{\parallel} =  \Omega_0\sum_{i=1}^{n} \sigma^x_j$ and $\hat{V}_{\#} =  \Omega_0\sum_{i=1}^{n} \prod_{j\ne i}^{n} \sigma^x_j$. Then
	\begin{eqnarray}
		\hspace{-1.9cm} v_{\parallel}^{\mathrm{dv}} &=& ||\hat{V}_{\parallel} - v^{\mathrm{min}}_{\parallel}|| =  \Omega_0 \,
		||\sum_{i} \sigma^x_j + n \mathds{1}|| = 
		2\,n \,\, \Omega_0 , \label{vdvp}\\
		\hspace{-1.9cm} v_{\#}^{\mathrm{dv}} &=& ||\hat{V}_{\#} - v^{\mathrm{min}}_{\#}|| =  
		\left\{
		\begin{array}{c}
			\Omega_0 \,
			||\sum_{i} \prod_{j\ne i} \sigma^x_j + n\mathds{1}|| = 2\,n\, \,\Omega_0 \,\,\,  \hspace{1.8cm}(\textrm{for n even}) ,\\
			\Omega_0 \,
			||\sum_{i} \prod_{j\ne i} \sigma^x_j + \left(n-2\right)\mathds{1}|| = 2\,(n-1)\, \,\Omega_0 \,\,\,  (\textrm{for n odd})  , 
		\end{array}
		\right.
		\label{vdvq}
	\end{eqnarray}
	where the last equality in Eqs.~(\ref{vdvp}) and~(\ref{vdvq}) are obtained by taking into account that all the individual operators in the sum commute among themselves and with the identity operator, exhibiting a 
	common basis of eigenvectors.    
	Hence, $r \leq 1$, indicating that the quantum charging considered here provides a genuine quantum charging advantage.
	
	
	\revisionRefA{Concerning decoherence effects, the QB performance will essentially depend on the physical system adopted. However, experimental implementations in superconducting quantum annealers could benefit from some intrinsic protection inherited from the  adiabatic dynamics. In fact, under Born-Markov approximation, Albash and Lidar have shown that, in a slowly-varying evolution through a time-dependent Hamiltonian~\cite{Albash:15},  decoherence effects on the system may occur in the \textit{instantaneous energy eigenbasis} of the Hamiltonian. This means that, apart from temperature effects, the adiabatic dynamics is protected by an evolution kept in the  ground state of the Hamiltonian. Naturally, this scenario may drastically change when temperature effects take relevant place, as the environment will push the system out from its lowest energy state.}

	\section{Conclusions}
	\label{section:conclusion}
	We have introduced a counter-diabatic approach to provide quantum extensive advantage for a QB with respect to classical batteries. In this scenario, we have shown that a spin-$1/2$ chain with just $O(n)$ terms of global Ising interactions is already capable of achieving quadratic instantaneous maximum power with respect to the number $n$ of particles, which is in contrast with the linear behavior for the classical charging. This method can be generalized for other physical systems for QBs. Indeed, the idea is to find out an adiabatic dynamics providing extensive advantage in terms of the normalized time $s$ via a tractable number of many-body interactions. Then, in order to remove the adiabatic time constraint, we can use a Floquet approximation for the counter-diabatic Hamiltonian so that fast charging can be kept without the necessity of adding extra many-body interactions. For future developments, we intend to consider more general battery models and physical implementations in experimental platforms available with current technology. 
	
	\section*{Acknowledgments}
	
	L.F.C.M. and A.C.D. acknowledge Coordena\c{c}\~ao de Aperfei\c{c}oamento de Pessoal de N\'{\i}vel Superior (CAPES) 
	for financial support. 
	M.S.S. is supported by Conselho Nacional de Desenvolvimento Cient\'{\i}fico e Tecnol\'ogico (CNPq) (307854/2020-5). 
	This research is also supported in part by CAPES (Finance Code 001) 
	and by the Brazilian National Institute for Science and Technology of Quantum Information (INCT-IQ).
	A. C. S acknowledges the support by the European Union's Horizon 2020 FET-Open project SuperQuLAN (899354), and by the Proyecto Sinérgico CAM 2020 Y2020/TCS-6545 (NanoQuCo-CM) from the Comunidad de Madrid.
	\section*{Bibliography}
	

\begin{thebibliography}{10}
		\expandafter\ifx\csname url\endcsname\relax
		\def\url#1{{\tt #1}}\fi
		\expandafter\ifx\csname urlprefix\endcsname\relax\def\urlprefix{URL }\fi
		\providecommand{\eprint}[2][]{\url{#2}}
		
		\bibitem{Alicki:13}
		Alicki R and Fannes M 2013 {\em Phys. Rev. E\/} {\bf 87}(4) 042123
		\urlprefix\url{https://link.aps.org/doi/10.1103/PhysRevE.87.042123}
		
		\bibitem{PRL2013Huber}
		Hovhannisyan K~V, Perarnau-Llobet M, Huber M and Ac\'{\i}n A 2013 {\em Phys.
			Rev. Lett.\/} {\bf 111} 240401
		\urlprefix\url{https://link.aps.org/doi/10.1103/PhysRevLett.111.240401}
		
		\bibitem{Campaioli:Book}
		Campaioli F, Pollock F~A and Vinjanampathy S 2018 {\em Quantum Batteries\/}
		(Cham: Springer International Publishing) pp 207--225 ISBN 978-3-319-99046-0
		\urlprefix\url{https://doi.org/10.1007/978-3-319-99046-0_8}
		
		\bibitem{PRL_Andolina}
		Andolina G~M, Keck M, Mari A, Campisi M, Giovannetti V and Polini M 2019 {\em
			Phys. Rev. Lett.\/} {\bf 122}(4) 047702
		\urlprefix\url{https://link.aps.org/doi/10.1103/PhysRevLett.122.047702}
		
		\bibitem{kieu:04}
		Kieu T~D 2004 {\em Phys. Rev. Lett.\/} {\bf 93} 140403
		\urlprefix\url{https://journals.aps.org/prl/abstract/10.1103/PhysRevLett.93.140403}
		
		\bibitem{Bhattacharjee:21}
		Bhattacharjee S and Dutta A 2021 {\em Eur. Phys. J. B\/} {\bf 94} 239
		\urlprefix\url{https://link.springer.com/article/10.1140/epjb/s10051-021-00235-3}
		
		\bibitem{Campaioli:23}
		Campaioli F, Gherardini S, Quach J~Q, Polini M and Andolina G~M 2024 {\em Rev.
			Mod. Phys.\/} {\bf 96}(3) 031001
		\urlprefix\url{https://link.aps.org/doi/10.1103/RevModPhys.96.031001}
		
		\bibitem{PRB2019Batteries}
		Farina D, Andolina G~M, Mari A, Polini M and Giovannetti V 2019 {\em Phys. Rev.
			B\/} {\bf 99} 035421
		\urlprefix\url{https://link.aps.org/doi/10.1103/PhysRevB.99.035421}
		
		\bibitem{Carrega:20}
		Carrega M, Crescente A, Ferraro D and Sassetti M 2020 {\em New J. Phys.\/} {\bf
			22} 083085
		\urlprefix\url{https://iopscience.iop.org/article/10.1088/1367-2630/abaa01}
		
		\bibitem{Mojaveri:23}
		Mojaveri B, Jafarzadeh~Bahrbeig R, Fasihi M~A and Babanzadeh S 2023 {\em Sci.
			Rep.\/} {\bf 13} 19827
		\urlprefix\url{https://www.nature.com/articles/s41598-023-47193-7}
		
		\bibitem{Arjmandi:23}
		Arjmandi M~B, Mohammadi H, Saguia A, Sarandy M~S and Santos A~C 2023 {\em Phys.
			Rev. E\/} {\bf 108} 064106
		\urlprefix\url{https://journals.aps.org/pre/abstract/10.1103/PhysRevE.108.064106}
		
		\bibitem{Lu:24}
		{Lu} Z~G, {Tian} G, {L{\"u}} X~Y and {Shang} C 2024 {\em arXiv e-prints\/}
		arXiv:2405.03675 (\textit{Preprint} \eprint{2405.03675})
		\urlprefix\url{https://arxiv.org/abs/2405.03675}
		
		\bibitem{Ahmadi:24}
		Ahmadi B, Mazurek P, Horodecki P and Barzanjeh S 2024 {\em Phys. Rev. Lett.\/}
		{\bf 132}(21) 210402
		\urlprefix\url{https://link.aps.org/doi/10.1103/PhysRevLett.132.210402}
		
		\bibitem{Santos:19-a}
		Santos A~C, \c{C}akmak B, Campbell S and Zinner N~T 2019 {\em Phys. Rev. E\/}
		{\bf 100}(3) 032107
		\urlprefix\url{https://link.aps.org/doi/10.1103/PhysRevE.100.032107}
		
		\bibitem{Santos:20PRE}
		Santos A~C, Saguia A and Sarandy M~S 2020 {\em Phys. Rev. E\/} {\bf 101}(6)
		062114 \urlprefix\url{https://link.aps.org/doi/10.1103/PhysRevE.101.062114}
		
		\bibitem{Moraes:21}
		Moraes L~F~C, Saguia A, Santos A~C and Sarandy M~S 2021 {\em EPL (Europhys.
			Lett.)\/} {\bf 136} 23001
		\urlprefix\url{https://iopscience.iop.org/article/10.1209/0295-5075/ac1363}
		
		\bibitem{Abel:24}
		{Rojo-Franc{\`a}s} A, {Isaule} F, {Santos} A~C, {Juli{\'a}-D{\'\i}az} B and
		{Zinner} N~T 2024 {\em arXiv e-prints\/} arXiv:2406.07397 (\textit{Preprint}
		\eprint{2406.07397})
		
		\bibitem{gyhm_beneficial_2024}
		Gyhm J~Y and Fischer U~R 2024 {\em AVS Quantum Science\/} {\bf 6} 012001 ISSN
		2639-0213 \urlprefix\url{https://doi.org/10.1116/5.0184903}
		
		\bibitem{Celeri:17PRL}
		Campaioli F, Pollock F~A, Binder F~C, C\'eleri L, Goold J, Vinjanampathy S and
		Modi K 2017 {\em Phys. Rev. Lett.\/} {\bf 118}(15) 150601
		\urlprefix\url{https://link.aps.org/doi/10.1103/PhysRevLett.118.150601}
		
		\bibitem{Rossini:20-PRL}
		Rossini D, Andolina G~M, Rosa D, Carrega M and Polini M 2020 {\em Phys. Rev.
			Lett.\/} {\bf 125}(23) 236402
		\urlprefix\url{https://link.aps.org/doi/10.1103/PhysRevLett.125.236402}
		
		\bibitem{Gyhm:22}
		Gyhm J~Y, \ifmmode~\check{S}\else \v{S}\fi{}afr\'anek D and Rosa D 2022 {\em
			Phys. Rev. Lett.\/} {\bf 128}(14) 140501
		\urlprefix\url{https://link.aps.org/doi/10.1103/PhysRevLett.128.140501}
		
		\bibitem{Roland:02}
		Roland J and Cerf N~J 2002 {\em Phys. Rev. A\/} {\bf 65}(4) 042308
		\urlprefix\url{https://link.aps.org/doi/10.1103/PhysRevA.65.042308}
		
		\bibitem{Le:18}
		Le T~P, Levinsen J, Modi K, Parish M~M and Pollock F~A 2018 {\em Phys. Rev.
			A\/} {\bf 97}(2) 022106
		\urlprefix\url{https://link.aps.org/doi/10.1103/PhysRevA.97.022106}
		
		\bibitem{Barra_2022}
		Barra F, Hovhannisyan K~V and Imparato A 2022 {\em New Journal of Physics\/}
		{\bf 24} 015003 \urlprefix\url{https://dx.doi.org/10.1088/1367-2630/ac43ed}
		
		\bibitem{Grazi:24}
		{Grazi} R, {Sacco Shaikh} D, {Sassetti} M, {Traverso Ziani} N and {Ferraro} D
		2024 {\em arXiv e-prints\/} arXiv:2402.09169
		\urlprefix\url{https://arxiv.org/abs/2402.09169}
		
		\bibitem{Salvia:23}
		Salvia R, Perarnau-Llobet M, Haack G, Brunner N and Nimmrichter S 2023 {\em
			Phys. Rev. Res.\/} {\bf 5}(1) 013155
		\urlprefix\url{https://link.aps.org/doi/10.1103/PhysRevResearch.5.013155}
		
		\bibitem{Santos:21}
		Santos A~C 2021 {\em Phys. Rev. E\/} {\bf 103}(4) 042118
		\urlprefix\url{https://link.aps.org/doi/10.1103/PhysRevE.103.042118}
		
		\bibitem{Dou:22}
		Dou F~Q, Zhou H and Sun J~A 2022 {\em Phys. Rev. A\/} {\bf 106}(3) 032212
		\urlprefix\url{https://link.aps.org/doi/10.1103/PhysRevA.106.032212}
		
		\bibitem{Demirplak:03}
		Demirplak M and Rice S~A 2003 {\em J. Phys. Chem. A\/} {\bf 107} 9937
		\urlprefix\url{https://doi.org/10.1021/jp030708a}
		
		\bibitem{Demirplak:05}
		Demirplak M and Rice S~A 2005 {\em J. Phys. Chem. B\/} {\bf 109} 6838
		\urlprefix\url{https://doi.org/10.1021/jp040647w}
		
		\bibitem{Berry:09}
		Berry M 2009 {\em J. Phys. A: Math. Theor.\/} {\bf 42} 365303
		\urlprefix\url{https://doi.org/10.1088%2F1751-8113%2F42%2F36%2F365303}
		
		\bibitem{Torrontegui:13}
		Torrontegui E, Ib{\'a}{\~n}ez S, Mart{\'\i}nez-Garaot S, Modugno M, Del~Campo
		A, Gu{\'e}ry-Odelin D, Ruschhaupt A, Chen X and Muga J~G 2013 {\em Adv. At.
			Mol. Opt. Phys\/} {\bf 62} 117
		\urlprefix\url{https://doi.org/10.1103/RevModPhys.91.045001}
		
		\bibitem{KOLODRUBETZ20171}
		Kolodrubetz M, Sels D, Mehta P and Polkovnikov A 2017 {\em Physics Reports\/}
		{\bf 697} 1--87 ISSN 0370-1573 geometry and non-adiabatic response in quantum
		and classical systems
		\urlprefix\url{https://www.sciencedirect.com/science/article/pii/S0370157317301989}
		
		\bibitem{Claeys:19}
		Claeys P~W, Pandey M, Sels D and Polkovnikov A 2019 {\em Phys. Rev. Lett.\/}
		{\bf 123}(9) 090602
		\urlprefix\url{https://link.aps.org/doi/10.1103/PhysRevLett.123.090602}
		
		\bibitem{Allahverdyan:04}
		Allahverdyan A~E, Balian R and Nieuwenhuizen T~M 2004 {\em Europhys. Lett.\/}
		{\bf 67} 565
		\urlprefix\url{https://iopscience.iop.org/article/10.1209/epl/i2004-10101-2}
		
		\bibitem{Coulamy:17}
		Coulamy I~B, Saguia A and Sarandy M~S 2017 {\em Phys. Rev. E\/} {\bf 95}(2)
		022127 \urlprefix\url{https://link.aps.org/doi/10.1103/PhysRevE.95.022127}
		
		\bibitem{Rezakhani:09}
		Rezakhani A~T, Kuo W~J, Hamma A, Lidar D~A and Zanardi P 2009 {\em Phys. Rev.
			Lett.\/} {\bf 103}(8) 080502
		\urlprefix\url{https://link.aps.org/doi/10.1103/PhysRevLett.103.080502}
		
		\bibitem{Hu:22a}
		Hu C~K, Qiu J, Souza P~J~P, Yuan J, Zhou Y, Zhang L, Chu J, Pan X, Hu L, Li J,
		Xu Y, Zhong Y, Liu S, Yan F, Tan D, Bachelard R, Villas-Boas C~J, Santos A~C
		and Yu D 2022 {\em Quantum Science and Technology\/} {\bf 7} 045018
		\urlprefix\url{https://doi.org/10.1088/2058-9565/ac8444}
		
		\bibitem{Rezakhani:10}
		Rezakhani A~T, Pimachev A~K and Lidar D~A 2010 {\em Phys. Rev. A\/} {\bf 82}(5)
		052305 \urlprefix\url{https://link.aps.org/doi/10.1103/PhysRevA.82.052305}
		
		\bibitem{Hu-Biao:16}
		Hu H and Wu B 2016 {\em Phys. Rev. A\/} {\bf 93}(1) 012345
		\urlprefix\url{https://link.aps.org/doi/10.1103/PhysRevA.93.012345}
		
		\bibitem{Messiah:Book}
		Messiah A 1962 {\em Quantum Mechanics\/} Quantum Mechanics (North-Holland
		Publishing Company) ISBN 9780720400441
		
		\bibitem{Leibfried:03}
		Leibfried D, Blatt R, Monroe C and Wineland D 2003 {\em Rev. Mod. Phys.\/} {\bf
			75}(1) 281--324
		\urlprefix\url{https://link.aps.org/doi/10.1103/RevModPhys.75.281}
		
		\bibitem{wendin2017}
		Wendin G 2017 {\em Reports on Progress in Physics\/} {\bf 80} 106001
		\urlprefix\url{https://doi.org/10.1088%2F1361-6633%2Faa7e1a}
		
		\bibitem{Rasmussen:21}
		Rasmussen S, Christensen K, Pedersen S, Kristensen L, B\ae{}kkegaard T, Loft N
		and Zinner N 2021 {\em PRX Quantum\/} {\bf 2}(4) 040204
		\urlprefix\url{https://link.aps.org/doi/10.1103/PRXQuantum.2.040204}
		
		\bibitem{Hu:23}
		Hu C~K, Yuan J, Veloso B~A, Qiu J, Zhou Y, Zhang L, Chu J, Nurbolat O, Hu L, Li
		J, Xu Y, Zhong Y, Liu S, Yan F, Tan D, Bachelard R, Santos A~C, Villas-Boas C
		and Yu D 2023 {\em Phys. Rev. Appl.\/} {\bf 20}(3) 034072
		\urlprefix\url{https://link.aps.org/doi/10.1103/PhysRevApplied.20.034072}
		
		\bibitem{Katz:23}
		Katz O, Cetina M and Monroe C 2023 {\em PRX Quantum\/} {\bf 4}(3) 030311
		\urlprefix\url{https://link.aps.org/doi/10.1103/PRXQuantum.4.030311}
		
		\bibitem{Zhang_SQC:22}
		Zhang K, Li H, Zhang P, Yuan J, Chen J, Ren W, Wang Z, Song C, Wang D~W, Wang
		H, Zhu S, Agarwal G~S and Scully M~O 2022 {\em Phys. Rev. Lett.\/} {\bf
			128}(19) 190502
		\urlprefix\url{https://link.aps.org/doi/10.1103/PhysRevLett.128.190502}
		
		\bibitem{Albash:15}
		Albash T and Lidar D~A 2015 {\em Phys. Rev. A\/} {\bf 91}(6) 062320
		\urlprefix\url{https://link.aps.org/doi/10.1103/PhysRevA.91.062320}
		
	\end{thebibliography}
	
	\providecommand{\newblock}{}

\end{document}